# MULTISCALE MODELING OF PLASTIC DEFORMATION OF MOLYBDENUM AND TUNGSTEN:
# I. ATOMISTIC STUDIES OF THE CORE STRUCTURE AND GLIDE OF 1/2⟨111⟩ SCREW DISLOCATIONS AT 0 K

R. Gröger[1,2]*, A. G. Bailey[1,3], V. Vitek[1]

[1] University of Pennsylvania, Department of Materials Science and Engineering, Philadelphia, PA 19104, USA
[2] Los Alamos National Laboratory, Theoretical Division, Los Alamos, NM 87545, USA
[3] Imperial College, Physics Department, London SW7 2AZ, UK

*Corresponding author. *E-mail:* groger@lanl.gov

**Abstract**
Owing to their non-planar cores 1/2⟨111⟩ screw dislocations govern the plastic deformation of BCC metals. Atomistic studies of the glide of these dislocations at 0 K have been performed using Bond Order Potentials for molybdenum and tungsten that account for the mixed metallic and covalent bonding in transition metals. When applying pure shear stress in the slip direction it displays significant twinning-antitwinning asymmetry for molybdenum but not for tungsten. However, for tensile/compressive loading the Schmid law breaks down in both metals, principally due to the effect of shear stresses perpendicular to the slip direction that alter the dislocation core. Recognition of this phenomenon forms a basis for the development of physically based yield criteria that capture the breakdown of the Schmid law in BCC metals. Moreover, dislocation glide may be preferred on {110} planes other than the most highly stressed one, which is reminiscent of the anomalous slip observed in many BCC metals.

*Keywords:* atomistic modeling; dislocations; Peierls stress; Schmid law; Bond Order Potential; molybdenum; tungsten.

## 1. Introduction

The most prominent features of the plastic deformation of body-centered cubic (BCC) metals are the rapid increase of the flow stress with decreasing temperature and increasing strain rate. In addition, in single crystals the yield and flow stress depend strongly on the orientation of the crystal relative to the loading axes and the Schmid law does not apply (for reviews see [1-6]). These aspects of yielding have been found in a broad variety of non-magnetic transition metals of the VB and VIB groups of the periodic table (e.g. [7-20], in iron and iron-silicon alloys [21-28] as well as in alkali metals [29-34]. Niobium, molybdenum and tantalum are the transition metals that were studied most extensively [7-20] while tungsten was investigated much less frequently. The main reason is that its purification is rather involved and preparation of single crystals is difficult owing to its extremely high melting temperature (3422°C) and high brittle-to-ductile transition temperature [35, 36]. Up to date there has been only a handful of experimental studies of the plastic deformation of tungsten single crystals [37-39] and these were usually made at room temperature while in other refractory metals the measurements commonly extend to 77 K or even lower temperatures.

It has been firmly established by many experimental and theoretical studies performed in the last forty years that the strong temperature and strain rate dependence of the flow stress results from the high lattice friction (Peierls) stress of 1/2⟨111⟩ screw dislocations. As first suggested by Hirsch [40], the reason is that their cores spread into several planes of the ⟨111⟩ zone, which is now a generally accepted characteristic of screw dislocations in BCC crystals. This type of dislocation core is then responsible not only for the high Peierls stress but also for all the other outstanding features of the plastic deformation of BCC metals [41]. While direct experimental observations of the cores of screw dislocations are precarious [42, 43], such core spreading has been found in all atomistic calculations (for reviews see [3, 44-51]). In all these studies the core was found to spread into three {110} planes, specifically $(\bar{1}01)$, $(0\bar{1}1)$ and $(\bar{1}10)$ for the 1/2[111] dislocation.

The main factor controlling the possible core spreading is, as in the case of all crystal defects, the symmetry of the crystal structure. Following the so-called Neumann's principle [52], symmetry of any physical property has to comprise the symmetry operations of the underlying point group of the crystal lattice. This does not imply, of course, that the structure of every crystal defect must be invariant to all symmetry operations associated with a given structure. Yet, if the structure of a defect is not invariant with respect to a symmetry operation, degenerate (energetically equivalent) configurations exist that are related by this symmetry operation. The most important symmetry relevant for the 1/2[111] screw dislocation in a BCC crystal is that [111] is the direction of a three-fold screw axis. This was first noted by Suzuki [53] who argued that 1/2⟨111⟩ screw dislocations must possess such symmetry and are thus intrinsically non-planar. Indeed, atomistic calculations that employed a wide variety of descriptions of interatomic interactions all confirmed this symmetry of the core. However, another symmetry operation pertinent to line defects parallel to the [111] direction is the $[\bar{1}01]$ diad, which can also be considered as the reflection in the (111) plane followed by the reflection in the $(\bar{1}2\bar{1})$ plane. The core structure of the 1/2[111] screw dislocation obeying this symmetry was found in recent calculations employing DFT-based methods [54-58], tight-binding based Bond Order Potentials [59, 60] and also some central force potentials [45, 61-64]; this is the core structure of screw dislocations analyzed in the present paper. However, most of the previous calculations led to core structures that are not invariant with respect to the $[\bar{1}01]$ diad (see reviews [44, 47, 49, 65, 66]). In this case there are two energetically equivalent configurations related by the operation of the $[\bar{1}01]$ diad[1]. Consequently, which of the core structures is found depends on the description of atomic interactions used in the calculations.

In transition metals the bonding has a mixed nearly free-electron and covalent character and it is the filling of the d-band that controls the cohesion [67, 68]. This bonding, which is mediated by the d-electrons, leads to the atomic interactions that are non-central and depend on the angles between bonds. Hence, central force potentials cannot reflect with sufficient accuracy the bonding aspects specific to a given transition metal although atomistic calculations employing such potentials can reveal general characteristics of the lattice defects related to the BCC structure. The physics of bonding is reflected most reliably in the methods based on the density functional theory, which are at present state-of-the-art. They have, indeed, been employed in investigations of mechanical properties of materials (for reviews see [69, 70]), specifically in studies of screw dislocations in molybdenum and tantalum [54-56, 71]. However, these calculations are still limited by feasible block sizes and by the use of periodic boundary conditions. The latter are particularly troublesome when studying the motion of dislocations under the effect of applied stresses. Consequently, approximations and simplifications are needed when describing interatomic interactions that must not, however, obliterate essential features of the bonding, such as correct directional character of bonds. One such method is the Bond Order Potential (BOP) introduced by Pettifor and co-workers [72-75]. This method is particularly suitable for modeling of extended defects since it can be utilized in real space and thus periodic boundary conditions are not needed. It is based on chemically intuitive tight-binding approximation to the quantum mechanical electronic structure with the electronic degrees of freedom coarse-grained into a many-body interatomic potential that reflects correctly the angular character of bonding (for reviews see [76-81]). Recently, we have developed BOPs for both molybdenum [59] and tungsten [60]. These potentials were tested exhaustively for transferability to atomic environments that deviate very significantly from those of ideal lattice. Hence, they are likely to reveal correctly the structure and properties of lattice defects and all calculations presented in this paper have been made using these BOPs.

---

[1] In the [111] projection, which is used when depicting the screw dislocation cores, the structure invariant to the $[\bar{1}01]$ diad appears as six-fold and the non-invariant structure as three-fold. This terminology has often been used in the literature although it is not correct crystallographically.

In this paper we present results of atomistic calculations that focus on the complex response of 1/2[111] screw dislocations in molybdenum and tungsten at 0 K to externally applied stresses. Some preliminary results for molybdenum were published in [82, 83]. We start by calculating γ-surfaces that represent energies of generalized stacking faults formed by displacing two parts of a crystal relative to each other along a cut typically identified with a low-index crystallographic plane [44, 65, 84, 85]. These surfaces disclose any possible metastable stacking faults and relate to the dislocation core structure so that its main features can be anticipated from their form [45, 50, 61, 86]. As the next step, starting with a fully relaxed screw dislocation core, we investigate the effect of pure shear stress acting parallel to the Burgers vector. This study involves calculation of the critical resolved shear stress (CRSS) at which the dislocation starts to move for various orientations of the maximum resolved shear stress plane (MRSSP). The results demonstrate the dependence of the CRSS on the sense of shearing, reveal the so-called twinning-antitwinning asymmetry [3, 44] and illustrate the dependence of the CRSS on the orientation of the MRSSP that does not obey the Schmid law.

We then proceed to study the motion of screw dislocations under the effect of tensile and compressive stresses, the loading most frequently used in experimental studies. These calculations have been made for a number of differently oriented tension/compression axes and demonstrate that differences in shear stresses parallel to the Burgers vector are not capable to explain the variation of the CRSS with the orientation of the loading axis. This implies that other components of the stress tensor also affect the dislocation glide. Following the study of the influence of non-glide stresses on the motion of dislocations in BCC metals by Duesbery [87] and Ito and Vitek [61] we focus on shear stresses perpendicular to the Burgers vector. The ensuing dependence of the CRSS on these shear stresses explains the dependence of the CRSS on the orientation of the loading axis during tensile/compressive loading. Hence, glide of the 1/2[111] screw dislocation depends on shear stresses both parallel and perpendicular to the Burgers vector that act not only in the slip plane but also in other {110} planes of the [111] zone.

## 2. γ-surface for {110} planes

The concept of a γ-surface was introduced by defining a generalized stacking fault as a planar defect formed when a crystal is cut along a chosen crystal plane and the upper part displaced with respect to the lower part by a vector **t**, parallel to the plane of the cut [44, 51, 65, 84, 85]. The energy of the generalized stacking fault, γ(**t**), can be evaluated when an appropriate description of atomic interactions is available. As the vector **t** spans the unit cell in the plane of the cut, γ(**t**) generates a surface, commonly called the γ-surface. In such calculations relaxations of atoms in the direction perpendicular to the plane of the cut are usually allowed but not relaxations parallel to the cut since they would remove the disregistry introduced by the vector **t**. In the following we call such γ-surface *relaxed*, while if the positions of atoms are not relaxed but fixed in their displaced positions, the γ-surface is called *unrelaxed*.

As will be shown below, the screw dislocation cores tend to spread into {110} planes of the corresponding ⟨111⟩ zone and thus we concentrate on the γ-surface for the ($\bar{1}$01) plane. Our calculations utilized the BOPs for Mo and W [59, 60] and the overall shape of the γ-surface was found to be the same as in many previous calculations (see reviews [44, 47, 49, 51, 65, 66]). In particular, there are no minima that would indicate existence of metastable stacking faults. The same was also found for the (11$\bar{2}$) γ-surface. This finding appears to be a general feature of BCC metals and it implies that in these metals dislocations cannot dissociate into partial dislocations as it is common in FCC metals. The [111] cross-sections of the calculated ($\bar{1}$01) γ-surfaces are shown in Fig. 1, where they are compared with those found in analogous DFT-based calculations of Frederiksen and Jacobsen [57]. An excellent agreement between the BOP and DFT-based calculations is apparent.

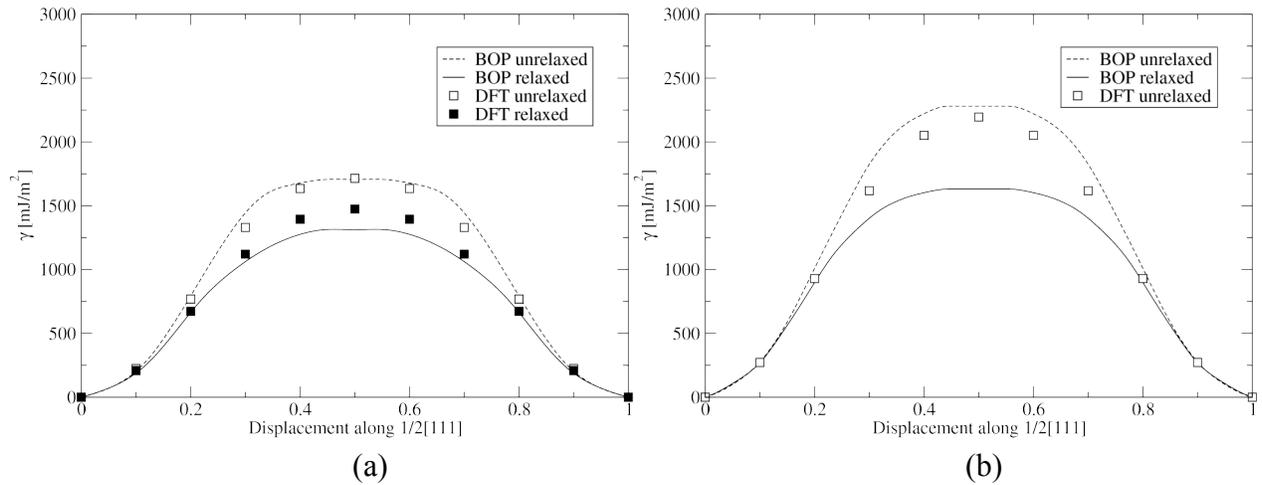

Fig. 1: [111] cross-sections of the ($\bar{1}$01) γ-surfaces for molybdenum (a) and tungsten (b) calculated using corresponding BOPs. The squares correspond to the DFT calculations of Frederiksen and Jacobsen [57].

## 3. Core structure of the 1/2[111] screw dislocation

In all the simulations presented in this paper we used a block of atoms that is oriented such that the *z*-axis coincides with the [111] direction (and thus with the Burgers vector and the direction of the dislocation line), *y*-axis is perpendicular to the $(\bar{1}01)$ plane, and *x* to both *y* and *z* such that the coordinate system is right-handed. To simulate an infinitely long straight screw dislocation, we use periodic boundary conditions along the *z*-direction. This reduces the number of (111) atomic layers in the model to three, with the nearest layer-to-layer distance equal to $1/2\sqrt{3}$ in units of the lattice parameter. The simulated block consisted of 1570 atoms and its extent in the *xy* plane was about $20 \times 20$ lattice parameters.

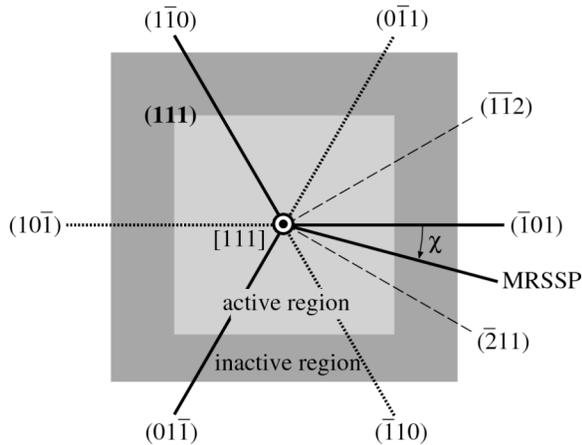

Fig. 2: Schematic picture of the block used in the atomistic calculations. The inactive region (dark) extends effectively to infinity. The orientation of the maximum resolved shear stress plane (MRSSP) is defined by the angle $\chi$ that it makes with the $(\bar{1}01)$ plane. Owing to the crystal symmetry, the two {112} planes, $(\bar{1}\bar{1}2)$ and $(\bar{2}11)$, represent the boundaries of the angular region of all orientations of the MRSSP that are not related by symmetry.

Starting with an ideal crystal, we inserted a 1/2[111] screw dislocation by displacing all the atoms in the block according to the anisotropic elastic strain field of the dislocation [88]. The atoms in the active region of the block (see Fig. 2) were subsequently relaxed using the corresponding BOP while holding the atoms in the inactive region fixed. The relaxation was considered complete when the forces on all atoms fell below 0.01 eV/Å. Since virtually the same core structure was found for both molybdenum and tungsten, Fig. 3a shows the relaxed core structure in both these metals using the usual differential displacement map [65, 85, 89]. The atomic arrangement is shown in the projection onto the (111) plane and the circles stand for atoms in the three successive (111) layers. The lengths of the arrows correspond to the displacements of two neighboring atoms parallel to the Burgers vector, i.e. perpendicular to the plane of the figure, relative to their distance in the perfect lattice. The three longest arrows close to the center of the figure, each corresponding to the relative displacement vector 1/6[111] in units of the lattice parameter, define a Burgers circuit that gives 1/2[111], the total Burgers vector of the dislocation. The same net product is obtained when going around the six second-largest arrows in the figure, each giving a relative displacement equal to 1/12[111], or around any other circuit encompassing the dislocation. This core structure is invariant with respect to both the [111] three-fold screw axis and to the $[\bar{1}01]$ diad and thus its structure is unique.

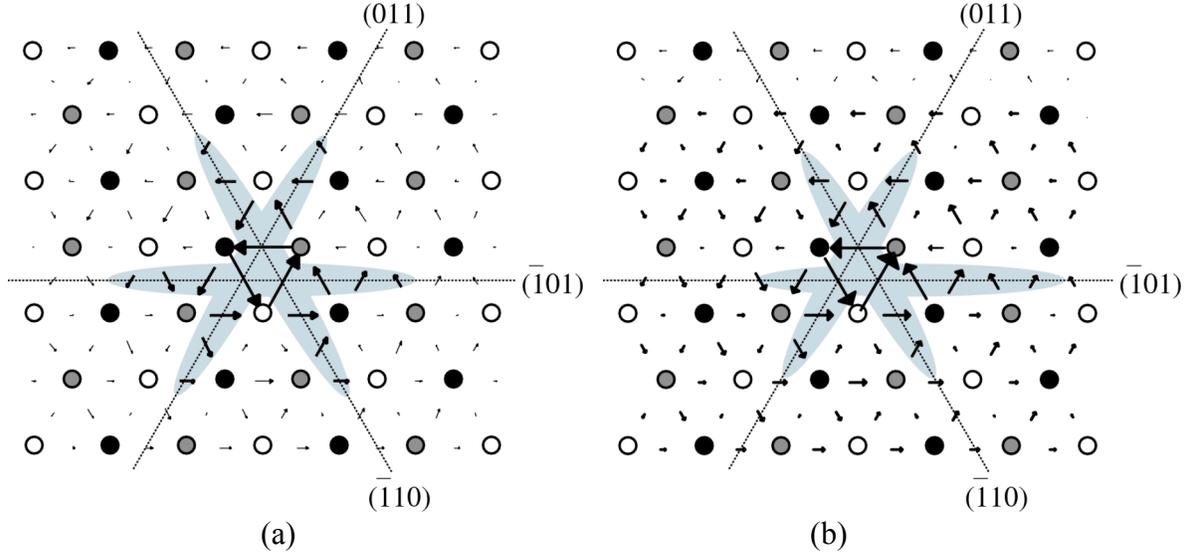

Fig. 3: (a) Structure of the 1/2[111] screw dislocation core relaxed using the BOP for molybdenum presented using the differential displacement map; the core structure found for tungsten is virtually identical. The circles represent positions of atoms in the three adjacent (111) planes, and the arrows are the displacements of two neighboring atoms in the [111] direction parallel to the Burgers vector (and thus perpendicular to the plane of the figure) relative to their separation in the perfect lattice. Shading is used to highlight the form of the core spreading. (b) Core of the 1/2[111] screw dislocation in molybdenum transformed under the effect of the shear stress $0.022C_{44}$ applied in the $(\bar{1}01)$ plane ($\chi = 0$). A similar core transformation occurs in tungsten.

## 4. Application of stress and glide of the screw dislocation

When applying the external stress the elastic displacement field corresponding to this stress was evaluated using anisotropic elasticity and superimposed on the dislocation displacement field for the atoms in both the active and inactive regions of the block. The atoms in the active region were then relaxed by minimizing the total energy of the block as in the non-stressed case. The goal of these calculations was to find the stress at which the dislocation starts to move, as well as the actual glide plane. For this purpose each calculation started with the fully relaxed unstressed core structure. The applied stress was increased incrementally and full relaxation carried out at every step until the dislocation started to move. This stress level was then identified with the critical stress needed for the dislocation glide at 0 K.

### *4.1 Loading by pure shear stress parallel to the slip direction*
According to the Schmid law, glide on a given slip system commences when the resolved shear stress on the slip plane in the direction of the slip reaches a critical value [90]. Implicitly, the sense of shearing and other components of the stress tensor than the shear stress in the slip plane parallel to the slip direction are assumed to play no role in the deformation process. The pure shear stress parallel to the slip direction was always applied in a maximum resolved shear stress plane (MRSSP) the orientation of which was defined by the angle $\chi$ that it makes with the $(\bar{1}01)$ plane, as shown in Fig. 2. This representation of the orientation of the MRSSP has been commonly used in earlier theoretical and experimental studies (see e.g. [3, 21, 44]). The dislocation starts to move when the shear stress in this plane reaches a critical value called the critical resolved shear stress (CRSS). If the Schmid law applied the projection of the CRSS onto the $(\bar{1}01)$ glide plane would be the same for any orientation of the MRSSP. In order to investigate possible deviation from the Schmid law we carried out a series of calculations for different orientations of the MRSSP. Owing to the crystal symmetry it is sufficient to consider $-30° \leq \chi \leq +30°$ but it should be noted that orientations corresponding to positive and negative angles $\chi$ are not equivalent. This range of $\chi$ is bounded by two {112} planes that are twinning planes in BCC crystals. For $\chi < 0$ the nearest {112} plane, i.e. $(\bar{1}\bar{1}2)$, is sheared in the *twinning* sense while for $\chi > 0$ the nearest {112} plane, i.e. $(\bar{2}11)$, is sheared in the *antitwinning* sense.

In the right-handed coordinate system with the *y*-axis normal to the MRSSP, the *z*-axis parallel to the slip direction and the *x*-axis in the MRSSP, the shear stress $\sigma$ parallel to the slip direction in the MRSSP is applied by the stress tensor

$$\Sigma_\sigma = \begin{pmatrix} 0 & 0 & 0 \\ 0 & 0 & \sigma \\ 0 & \sigma & 0 \end{pmatrix} . \qquad (1)$$

The shear stress $\sigma$ was built up incrementally in steps of $0.001C_{44}$, where $C_{44}$ is the elastic modulus. At low stresses, the dislocation core transforms from its symmetric non-degenerate configuration to a less symmetric form. This is seen in Fig. 3b that shows the core structure transformed under the effect of the shear stress $0.022C_{44}$, when the MRSSP is the $(\bar{1}01)$ plane ($\chi = 0$). This transformation is purely elastic in the sense that the structure returns into its original configuration when the stress is removed. However, once the applied shear stress $\sigma$ attains the CRSS, the transformation is no more elastic and the dislocation moves through the crystal.

The dependence of the CRSS on the orientation of the MRSSP, i.e. angle $\chi$, found in this study, is plotted as circles in Fig. 4a for molybdenum and in Fig. 4b for tungsten[2]. For all orientations of the MRSSP, the dislocation moved on the $(\bar{1}01)$ plane that coincides here with the most highly stressed {110} plane of the [111] zone. If the Schmid law applied the CRSS would vary with $\chi$ as $\cos^{-1}\chi$; these dependencies are shown as dashed curves in Fig. 4. In the case of molybdenum the calculated CRSS vs $\chi$ dependence evidently deviates from the Schmid law. For $\chi > 0$ (shear in the antitwinning sense) the CRSS is always higher than the corresponding value for $\chi < 0$ (shear in the twinning sense). This is the well-known twinning-antitwinning asymmetry often observed in BCC metals and regarded as the major contribution to the breakdown of the Schmid law (see e.g. [3, 44]). However, no such asymmetry was found in tungsten. Nonetheless, it does not follow that the Schmid law is valid in tungsten since other stress components may also affect the CRSS, as revealed in the following study that deals with tensile and compressive loadings.

---

[2] The calculated values of the CRSS (Peierls stress) are about two to three times higher than the measured CRSS extrapolated to 0 K. For example in [17, 18], where the CRSS was measured for molybdenum, it is $0.006\mu$, where $\mu$ is the shear modulus. The same disparity between calculations and observations was found in other elemental transition metals [27, 91] as well as in alkali metals [30, 33]. This discrepancy has been recently elucidated in [92]. It is not a consequence of any inadequacies in atomistic calculations but it is related to the fact that in real situations the dislocations never move in isolation but as a part of a large group of dislocations produced by a source. Due to the mutual interaction between emitted dislocations, the group consisting of both non-screw and screw dislocations can move at an applied stress that is a factor of two to three lower than the stress needed for the glide of individual screw dislocations.

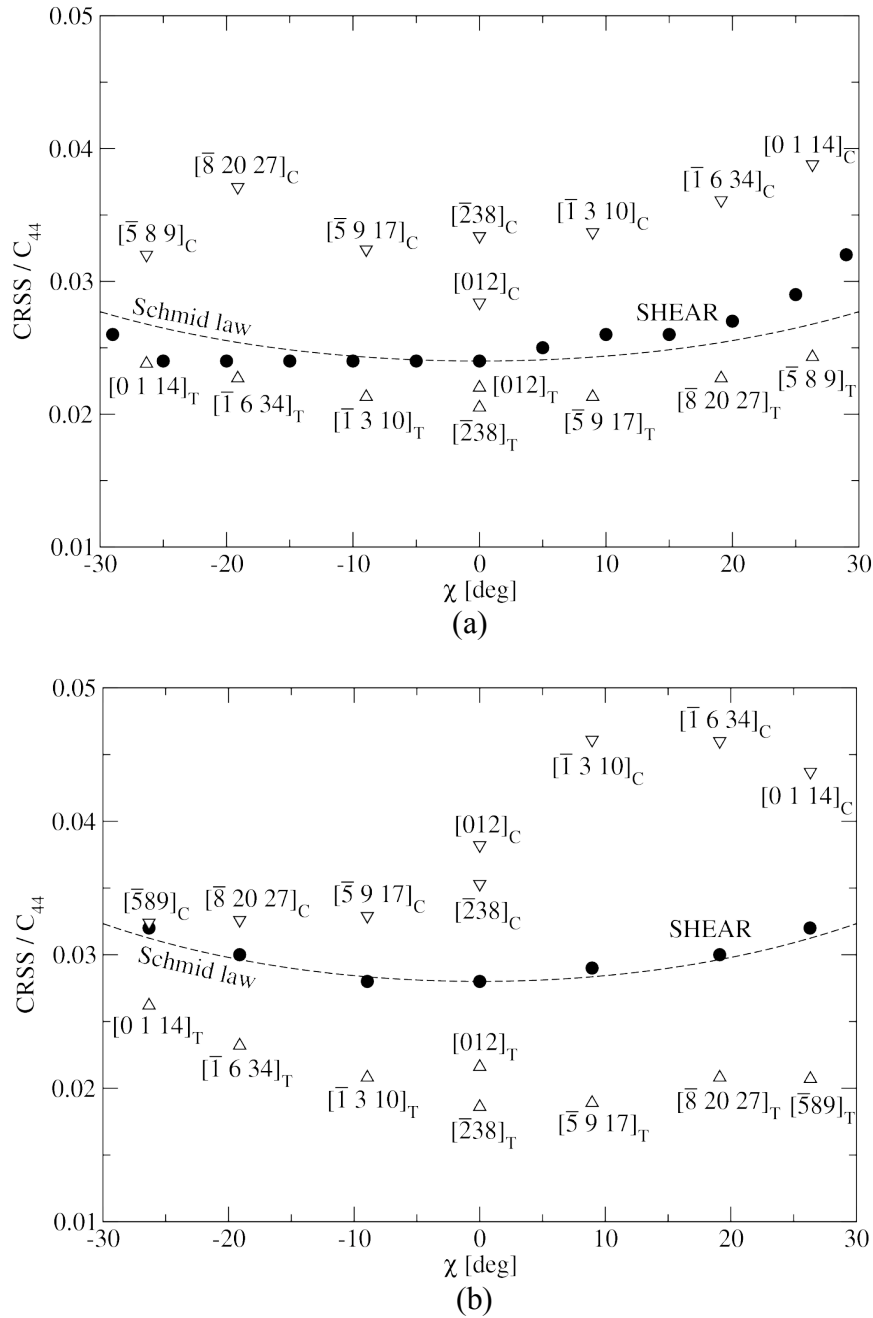

Fig. 4: Dependence of the CRSS on the orientation of the MRSSP ($\chi$) for loading by pure shear stress parallel to the slip direction in the MRSSP (circles), in tension (up-triangles) and compression (down-triangles); (a) molybdenum, (b) tungsten.

## 4.2 Loading in tension and compression

Simulation of the uniaxial tensile/compressive loading is an important test that can disclose whether or not the shear stress parallel to the slip direction is the only stress component that affects the dislocation movement. Such calculations need to be made for several different orientations of the loading axis and in each case the CRSS in the corresponding MRSSP evaluated. If for every loading axis this CRSS is the same as for loading by pure shear stress parallel to the slip direction in the corresponding MRSSP, we can conclude that no other stress component affects the CRSS at which the dislocation moves.

Owing to the symmetry of the BCC lattice, the complete set of non-equivalent loading axes can be identified within any stereographic triangle. We again consider that the $(\bar{1}01)$ plane is the most highly stressed {110} plane in the [111] zone and for this reason the stereographic triangle, the corners of which are identified by the axes [001], [011] and $[\bar{1}11]$, was chosen. It is shown in Fig. 5 together with the orientations of the tensile/compressive axes used in our calculations. The MRSSPs associated with individual loading axes are drawn in Fig. 5 as dashed curves that make various angles $\chi$ with the $(\bar{1}01)$ plane. The shear stress in these planes parallel to the [111] direction, at which the screw dislocation starts to move during loading in tension/compression, is the CRSS that can be directly compared with the CRSS found for the same angle $\chi$ when the pure shear stress parallel to the slip direction was applied. The tensile/compressive stress was again built up incrementally in steps of $0.001 C_{44}$ until the dislocation started to move.

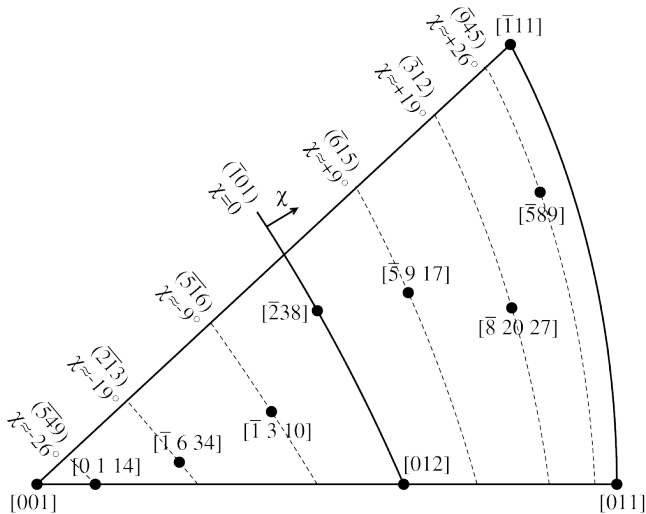

Fig. 5: Standard stereographic triangle in which the $(\bar{1}01)$ plane is the most highly stressed {110} plane in the zone of the [111] slip direction for any orientation of the tensile/compressive axis. Orientations of the loading axes are in square brackets and the corresponding MRSSPs together with the concurrent angles $\chi$ for loading in tension are marked along the [001]-$[\bar{1}11]$ side. The sign of $\chi$ changes for loading in compression.

The values of the CRSS found for loading in tension are drawn in Fig. 4 as up-triangles and for compression as down-triangles. It is seen from this figure that for a given $\chi$ the CRSS for tension, compression and pure shear in the slip direction differ considerably in both molybdenum and tungsten. This implies that in both metals the glide of the screw dislocations is also significantly affected by other stress components. The analysis of this type was performed originally by Ito and Vitek [61] in an atomistic study of 1/2[111] screw dislocations that employed central-force many body potentials of the Finnis-Sinclair type for molybdenum and tantalum [93, 94]. While these potentials do not reflect correctly the significant homeopolar component arising in transition metals owing to the partially filled d-band [67, 68], this study still uncovered which stress tensor components other than the shear stress in the slip direction may affect the glide of screw dislocations in BCC crystals. They are the shear stresses perpendicular to the slip direction. In the following section we investigate their effect on the dislocation glide, in particular their effect on the CRSS, and demonstrate that they are, indeed, the stress components responsible for differences between loadings in tension, compression and pure shear.

*4.3 Loading by shear stress perpendicular to the slip direction combined with shear stress parallel to the slip direction*

The stress tensor applied in this study has the form

$$\mathbf{\Sigma}_{\tau,\sigma} = \begin{pmatrix} -\tau & 0 & 0 \\ 0 & \tau & \sigma \\ 0 & \sigma & 0 \end{pmatrix} \qquad (2)$$

and relates to the right-handed coordinate system with the z-axis parallel to the dislocation line and the [111] slip direction, y-axis normal to the MRSSP in which the shear stress $\sigma$ parallel to the [111] direction is applied, and the x-axis in the MRSSP. The orientation of the MRSSP with respect to the $(\bar{1}01)$ plane is again defined by the angle $\chi$. $\tau$ is the magnitude of the shear stress perpendicular to the slip direction that is represented in this coordinate system as a combination of two normal stresses of opposite sign[3]. The stress tensor given by (2) cannot be applied experimentally but in computer simulations it allows us to differentiate the effects of the two types of shear stresses on the dislocation core. As will be described below, it can be linked with the studies of tension/compression loadings by determining for each loading axis the corresponding values of $\tau$ and $\sigma$ in the coordinate system defined above; the corresponding stress tensor will have, of course, also other non-zero components, e.g. hydrostatic stress, but these do not affect the dislocation core.

Prior to the study of the dislocation glide when applying the stress tensor (2), it is instructive to investigate the dislocation core changes induced solely by the shear stress perpendicular to the slip direction. This can be done by applying the stress tensor (2) with $\sigma = 0$. Since the shear stress perpendicular to the slip direction does not exert any Peach-Koehler force on the dislocation and, therefore, cannot produce any dislocation motion, the core changes induced by this stress are purely elastic, meaning that the structure returns to the unstressed state if the stress is removed. We present here only the results for $\chi = 0$, i.e. for the orientation when the y-axis coincides with the normal of the $(\bar{1}01)$ plane, but the same calculations were also made for other orientations. In these calculations, the stress $\tau$ was built up incrementally in steps of $0.005C_{44}$. The structure of the core of the 1/2[111] screw dislocation in molybdenum when $\tau = +0.05C_{44}$ is shown in Fig. 6a and for $\tau = -0.05C_{44}$ in Fig. 6b. The core structure in tungsten is not shown because it is practically the same as in molybdenum. For the positive $\tau$ the dislocation core extends further onto the $(\bar{1}01)$ plane while it constricts on both $(0\bar{1}1)$ and $(\bar{1}10)$ planes. On the other hand, for negative τ, the core constricts on the ($\bar{1}$01) plane and extends on (0$\bar{1}$1) and ($\bar{1}$10) planes.

---

[3] The representation in which $\tau$ appears as a shear stress is obtained by rotating the coordinate system by the angle $-45°$ around the z-axis.

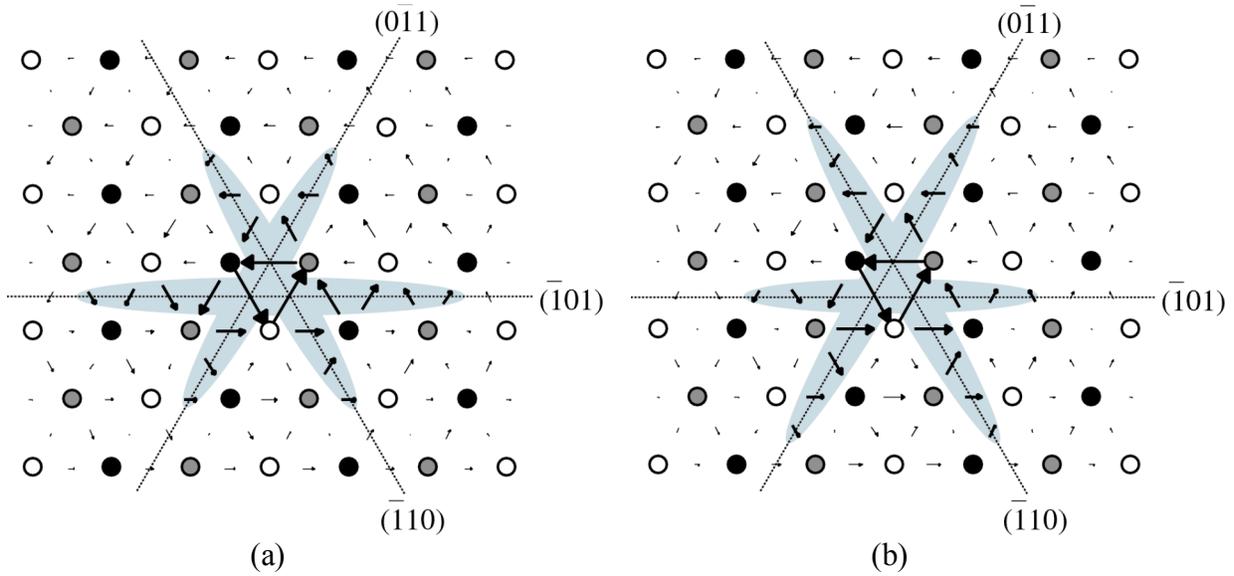

Fig. 6: Structure of the core of the 1/2[111] screw dislocation in molybdenum upon applying positive and negative shear stress perpendicular to the slip direction; (a) τ = +0.05$C_{44}$, (b) τ = –0.05$C_{44}$.

Based on this observation we can surmise that for positive τ the glide of the screw dislocation on the ($\bar{1}$01) plane, driven by the shear stress σ, will be easier when compared to the case of either none or negative shear stress perpendicular to the slip direction. Hence, one may expect that for τ > 0 the CRSS will be decreasing with increasing τ. Analogously, we can speculate that negative τ makes the glide on the ($\bar{1}$01) plane increasingly more difficult and at large negative τ the ($\bar{1}$01) glide may even be suppressed owing to the large extension of the core into the other two {110} planes of the [111] zone. Instead, the dislocation glide may proceed either on the (0$\bar{1}$1) or on the ($\bar{1}$10) plane even though the shear stress in the direction of the Burgers vector is in these planes smaller than in the ($\bar{1}$01) plane. In any case an appreciably larger CRSS can be expected at negative τ than when no shear stress perpendicular to the slip direction is applied. These conjectures have been corroborated by the calculations presented below.

The studies of the combined effect of the shear stresses parallel and perpendicular to the slip direction were performed for seven differently oriented MRSSPs that coincide with those that arise in the tension/compression calculations described earlier (see Fig. 5). In each case the shear stress perpendicular to the slip direction was first applied to reach a certain level of τ. This was done in steps as explained above. Keeping then the value of τ fixed, we built up incrementally the shear stress, σ, parallel to the slip direction until the CRSS, at which the dislocation moved into the next equivalent lattice site, was attained. These calculations yield the dependencies of the CRSS on the shear stress perpendicular to the slip direction for the MRSSPs studied. In Figs.

7 and 8 we show the CRSS vs $\tau$ dependencies for three typical orientations of the MRSSP for both molybdenum and tungsten. The MRSSPs are the $(\bar{1}01)$ plane corresponding to $\chi = 0$ and $(\bar{3}12)$ and $(\bar{2}\bar{1}3)$ planes corresponding, approximately, to $\chi = \pm 19°$, respectively. Results of calculations for other orientations of the MRSSPs ($\chi \approx \pm 9°, \pm 26°$), presented in [95], follow the same trends. In Figs. 7 and 8 the squares correspond to $\chi \leq 0$ and the circles to $\chi > 0$; the lines connecting the data points are included to lead the eye.

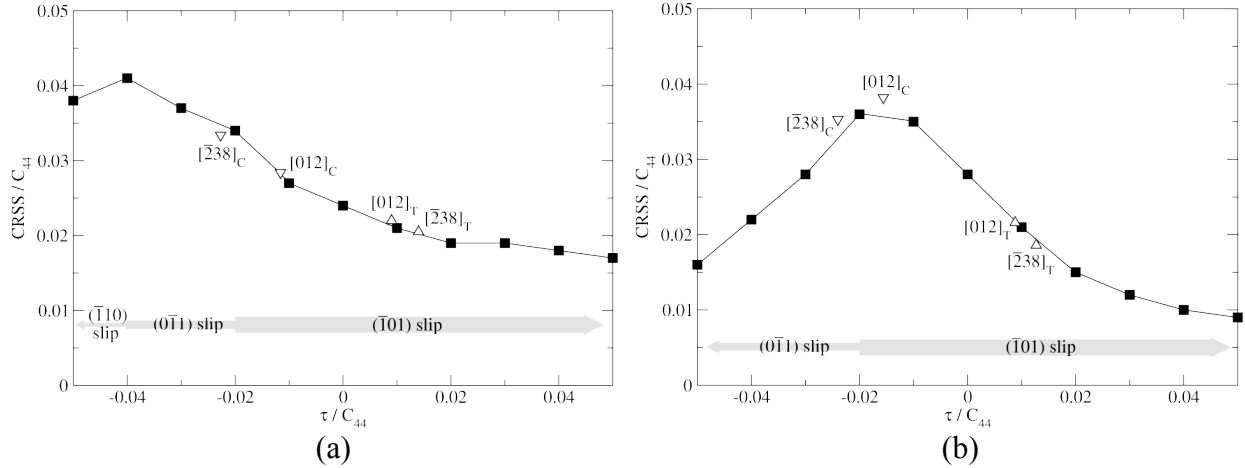

Fig. 7: Dependence of the CRSS on the shear stress perpendicular to the slip direction, $\tau$, when the MRSSP is the $(\bar{1}01)$ plane, i.e. $\chi = 0$; (a) molybdenum, (b) tungsten.

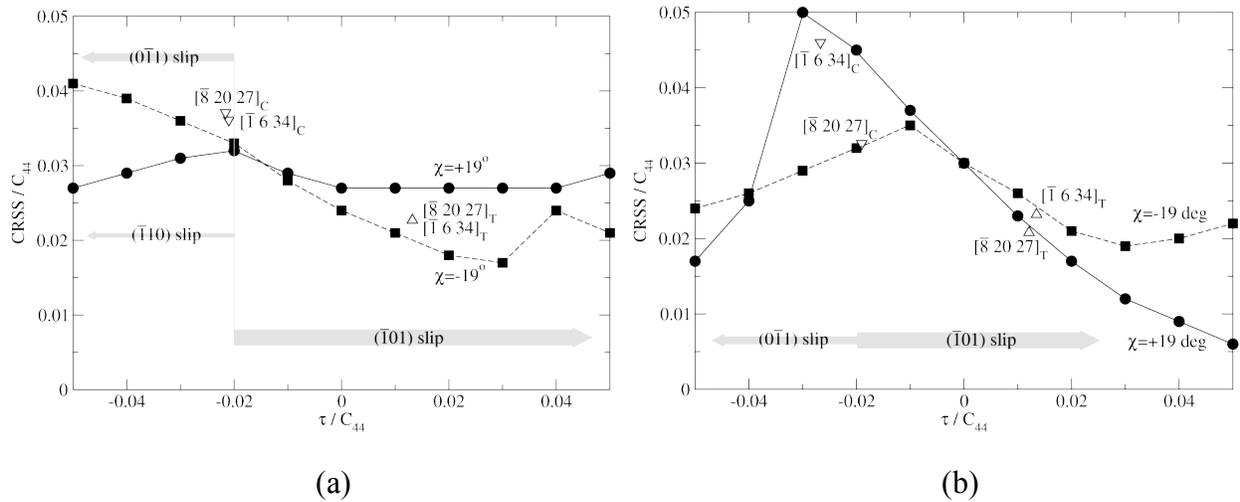

Fig. 8: Dependence of the CRSS on the shear stress perpendicular to the slip direction, $\tau$, for the MRSSPs $(\bar{3}12)$, $\chi = +19°$ and $(\bar{2}\bar{1}3)$, $\chi = -19°$; (a) molybdenum, (b) tungsten.

As we surmised, it is seen from Figs. 7 and 8 that for positive $\tau$ the CRSS is lower than for $\tau = 0$ and the dislocation always glides on the $(\bar{1}01)$ plane, just as when applying only the shear stress parallel to the slip direction. The reason is, as described above and shown in Fig. 6a, the dislocation core changes induced for $\tau > 0$. Thus the positive shear stress perpendicular to the slip direction promotes the dislocation glide on the $(\bar{1}01)$ plane by lowering the corresponding Peierls stress. In contrast, negative $\tau$ makes the slip on this plane more difficult. The reason is, as shown in Fig. 6b, that in this case the core is constricted on the $(\bar{1}01)$ plane and extended into $(0\bar{1}1)$ and $(\bar{1}10)$ planes. For $\tau > -0.02 C_{44}$, this core transformation is not yet big enough to alter the slip mode completely and the dislocation still glides on the $(\bar{1}01)$ plane. However, for larger negative values of $\tau$ the extension of the core into the $(0\bar{1}1)$ and $(\bar{1}10)$ planes becomes so overwhelming that the dislocation starts to glide on one of these planes. This occurs in spite of the fact that the Schmid factors corresponding to both the $(0\bar{1}1)[111]$ and $(\bar{1}10)[111]$ slip systems are a half of that for the most highly stressed $(\bar{1}01)[111]$ slip system.

Figs. 7 and 8 also comprise the values of CRSS and $\tau$, marked by triangles, that correspond to tension/compression with loading axes defined in Fig. 5. When evaluating $\tau$ and $\sigma$ for a given tensile/compressive axis only the deviatoric part of the stress tensor was used so that these values are directly comparable with those corresponding to the loading by a combination of the shear stresses perpendicular and parallel to the slip direction. $\tau$ is always positive for tension and it is negative for compression. A close agreement between the data for uniaxial loading and for loading by the shear stresses perpendicular and parallel to the slip direction is clearly visible. The same agreement was obtained for other orientations of the MRSSP that are not shown here but are presented in [95].

It is seen from Figs. 7 and 8 that while the general trend of the CRSS vs $\tau$ dependence is similar in both molybdenum and tungsten, the effect of the shear stress perpendicular to the slip direction is more significant in tungsten. The quantity that reflects the importance of the shear stress $\tau$ is the slope of the CRSS vs $\tau$ dependence for $\tau = 0$ that is much larger for tungsten than for molybdenum. As shown in Fig. 4b, there is virtually no twinning-antitwinning asymmetry in tungsten when loaded in pure shear. On the other hand, it is seen from the study of tensile/compressive loadings (Section 4.2) that the deviation from the Schmid law is larger in tungsten than in molybdenum. Hence, the breakdown of the Schmid law in tungsten is entirely due to the shear stresses perpendicular to the slip direction, while in molybdenum this is a consequence of both the effect of shear stresses perpendicular to the slip direction and the twinning-antitwinning asymmetry observed in pure shear parallel to the slip direction. This is, presumably, the reason why in tungsten the apparent CRSS for shearing in the twinning sense can be higher than for shearing in the antitwinning sense. Such situation is seen in Fig. 8b where for positive $\tau$ the CRSS is higher for $\chi = -19°$ than for $\chi = +19°$. Such situation does not arise in molybdenum (Fig. 8a) where for positive $\tau$ the CRSS is higher for $\chi = +19°$ than for $\chi = -19°$. In this case the twinning-antitwinning asymmetry related to shearing in the direction of slip is apparently significant. Notwithstanding, the effect of the shear stress perpendicular to the slip direction does contribute very significantly to the breakdown of the Schmid law in molybdenum but in tungsten it is the only contribution.

## 5. Conclusions and discussion

In this paper we have presented a comprehensive atomic-level study of the core structure and glide of $1/2\langle 111\rangle$ screw dislocations in molybdenum and tungsten. These calculations were made using a molecular statics method and, therefore, do not include any temperature effects. The interatomic interactions were described by the recently constructed Bond Order Potentials that reflect adequately the mixed nearly free-electron and covalent character of bonding in transition metals. Moreover, the transferability of these BOPs to environments that differ very significantly from the ideal BCC lattice was extensively tested and thus they are eminently suitable for atomistic studies of extended crystal defects [59, 60].

First of all, the calculations of γ-surfaces indicate that there are no metastable stacking faults on either {110} or {112} planes and thus no dislocation splitting into well-defined partials can occur. This is in agreement with all previous studies of BCC metals and it appears to be a general feature of materials with this crystal structure (see reviews [3, 44, 47, 51, 65]). In both metals studied, the core structure of 1/2[111] screw dislocations was found to be non-degenerate and thus unique, invariant with respect to both [111] three-fold screw axis and the [$\bar{1}$01] diad. This agrees with DFT-based calculations of the dislocation core in molybdenum [55, 56]. Furthermore, the recent DFT-based calculations of γ-surfaces for all group VB and VIB transition metals, as well as α-iron, suggest that in all these metals the cores of screw dislocations possess this symmetry [57]. This is in contrast to many previous studies in which degenerate dislocation cores that are not invariant with respect to the [$\bar{1}$01] diad were found. While the symmetry of the dislocation core is an important structural aspect, it is not determining uniquely the response of the core to an externally applied stress. In fact, the overall behavior of $1/2\langle 111\rangle$ screw dislocations under the effect of applied stresses is rather similar for the two types of core structures since, as discussed in more detail in [50], the symmetry related to the diad is eliminated by the stress application and the non-degenerate core attains a configuration akin to the degenerate one. Notwithstanding, peculiarities of the core structure specific to a given material, related to the details of bonding, are responsible for differences in dislocation glide behavior that may vary from one BCC metal to another. This is clearly demonstrated in the present study by different orientation dependencies of the critical resolved shear stress (CRSS) in molybdenum and tungsten while the unstressed dislocation cores appear to be practically identical in these two metals.

The result of the present atomistic studies that is common to both molybdenum and tungsten and to all modes of loading is that {110} planes are the glide planes of $1/2\langle 111\rangle$ screw dislocations at 0 K. This is in agreement with experimental observations of slip planes in molybdenum samples deformed at cryogenic temperatures, whether loaded in pure shear [14, 96, 97] or tension/compression [17, 98-102]. Moreover, {110} planes were found to be the principal slip planes in other BCC metals. Calculations for loading by pure shear stress parallel to the slip direction display the well-known twinning-antitwinning asymmetry for molybdenum. This asymmetry was, indeed, observed when single crystals of molybdenum were loaded by the shear stress at 77 K [14, 96, 97]. Moreover, it has been found in other BCC metals and regarded as the major reason for the breakdown of the Schmid law (see e.g. [3, 44]). Notwithstanding, in the present calculations such asymmetry was found to be very weak in tungsten.

In contrast to the twinning-antitwinning asymmetry, the breakdown of the Schmid law was found for both molybdenum and tungsten when simulating the tension/compression tests and it is, in fact, more pronounced for tungsten than for molybdenum. Explanation of this phenomenon has been provided by the calculations combining shear stresses parallel and perpendicular to the slip direction. These calculations show that the CRSS in the slip direction depends sensitively on the shear stress perpendicular to the slip direction. This confirms previous suggestions based on calculations employing central-force potentials [45, 61, 87, 103]. The reason is that the shear stress perpendicular to the slip direction changes the symmetry of the dislocation core such that it either promotes or impedes the slip on the most highly stressed {110} plane. In the right-handed coordinate system with the *z*-axis parallel to the [111] slip direction (and the dislocation line), *y*-axis normal to the MRSSP in which the shear stress parallel to the [111] direction is applied, and the *x*-axis in the MRSSP, the positive shear stress perpendicular to the slip direction lowers the CRSS while the CRSS increases when the shear stress perpendicular to the slip direction is negative. For the tension/compression axes within the stereographic triangle shown in Fig. 5, the shear stress perpendicular to the slip direction is positive for tension and negative for compression. This suggests that the CRSS in tension will be always lower than in compression, as is also seen in the present study of tensile/compressive loadings.

The investigation of the effect of the shear stresses perpendicular to the slip direction shows that in tungsten the breakdown of the Schmid law is entirely due to these shear stresses, whilst in molybdenum it is an outcome of combining the effect of these stresses with the twinning-antitwinning asymmetry. This can lead to an interesting situation, described in Section 4.3 (Fig. 8b) for the MRSSPs ($\bar{2}\bar{1}3$) and ($\bar{3}12$) ($\chi \approx \pm 19°$), respectively, that the CRSS may be higher for shearing in the twinning sense than in the antitwinning sense when the effect of the shear stress perpendicular to the slip direction dominates.

However, even more interesting is the finding that the dislocation may prefer to glide on {110} planes in which the resolved shear stress in the slip direction is lower than on the most highly stressed {110} plane. As described in Section 4.3, for large negative values of the shear stress perpendicular to the slip direction the 1/2[111] screw dislocation starts to glide on either ($0\bar{1}1$) or ($\bar{1}10$) planes although the Schmid factors corresponding to these planes are a half of that for the most highly stressed ($\bar{1}01$) plane. The reason is the contraction of the core in the latter plane and extension into the former planes. The occurrence of the slip on the ($0\bar{1}1$) and ($\bar{1}10$) planes is reminiscent of the *anomalous slip* that takes place on the slip systems for which the resolved shear stress is substantially lower than that for the most highly stressed {110}⟨111⟩ system. Such slip was observed in niobium [10, 11, 102, 104-107], tantalum [108], vanadium [109, 110] and molybdenum [100, 111, 112]. Indeed, the anomalous slip observed by Jeffcoat et al. [111] in Mo-5at%Nb and Mo-5at%Re deformed in compression at 77 K can be adequately explained on the basis of our results. In these experiments the orientation of the loading axis was given by the angle λ that it makes with the [111] direction, and by the angle χ between the MRSSP and the ($\bar{1}01$) plane. For λ = 50° and χ ≤ 0 the most prominent slip traces correspond to the ($0\bar{1}1$) plane. Besides this system, traces of the most highly stressed ($\bar{1}01$)[$\bar{1}\bar{1}\bar{1}$] system were also observed[4]. For these orientations of loading the shear stress perpendicular to the slip direction in the MRSSP is negative and following the above finding the $(0\bar{1}1)$ and $(\bar{1}10)$ slip planes may, indeed, be preferred over the $(\bar{1}01)$ plane that has the highest resolved shear stress in the slip direction.

Finally, the general inference based on the present atomistic study is that it is only the deviatoric part of the applied stress tensor that controls the glide of $1/2\langle 111\rangle$ screw dislocations. However, both the shear stress parallel to the slip direction whose critical value (CRSS) determines the onset of the dislocation motion and the shear stress perpendicular to the slip direction that changes the structure of the dislocation core but does not drive the dislocation impact significantly the dislocation glide. While this assertion has been proven only by comparing the tensile/compressive loading with the loading by combination of shear stresses parallel and perpendicular to the slip direction, it is likely to be also valid for other forms of stress application. Moreover, this finding is in agreement with Ref. [61] where central-force potentials were employed and the only orientations of the MRSSP studied corresponded to $\chi = 0°$ and $\chi = \pm 30°$. Hence, it is conceivable that it is valid not only in molybdenum and tungsten but also in other BCC metals.

Nevertheless, while these dependencies of the CRSS on the orientation of the MRSSP and on the shear stresses perpendicular to the slip direction are qualitatively similar in molybdenum and tungsten they do differ quantitatively. This demonstrates that they depend not only on crystallography but also on details of bonding in a given material. As proposed in [113-117], understanding of the role the whole applied stress tensor plays in the dislocation glide, revealed in atomistic studies like those presented in this paper, provides a basis for the formulation of yield criteria for continuum analyses of plasticity in both single crystals and polycrystals. This is the topic of the Part II of this series of papers where more detailed comparison with experimental observations is also presented.

**Acknowledgements**

This research was supported by the Department of Energy, BES Grant no. DE-PG02-98ER45702. During the course of this research we have benefited from extensive discussions with Prof. John Bassani and Dr. Vikranth Racherla (University of Pennsylvania). We also gratefully acknowledge the collaboration on development and use of Bond Order Potentials with Dr. Duc Nguyen-Manh (UKAEA, Culham Research Center) and Dr. Matous Mrovec (Fraunhofer Institut für Werkstoffmechanik, Germany).

---

[4] The slip planes observed in Ref. [111] were originally written with opposite slip directions. For consistency, we always use the Miller indices for which the shear stress parallel to the slip direction resolved in the MRSSP is positive. In compression, this means that instead of $(\bar{1}01)[111]$ we write $(\bar{1}01)[\bar{1}\bar{1}\bar{1}]$.